\documentclass[11pt]{article}
\usepackage[letterpaper, left=1in, top=1in, right=1in, bottom=1in,nohead,includefoot]{geometry}
\usepackage{color,charter,graphicx,natbib,here,setspace,multirow,amsmath,amssymb,amsfonts}
\usepackage{epstopdf}
\usepackage[dvipsnames]{xcolor}
\usepackage[compact]{titlesec}

\titlespacing{\section}{0pt}{5pt}{3pt}
\titlespacing{\subsection}{0pt}{5pt}{3pt}
\titlespacing{\subsubsection}{0pt}{4pt}{3pt}
\bibpunct{(}{)}{;}{a}{}{,}

	\usepackage[pdftex]{hyperref}
	\hypersetup{colorlinks,
	citecolor=NavyBlue  
	}


\begin{document}

\title{Volatility Forecasts Using Nonlinear Leverage Effects}
\author{
Kenichiro McAlinn\thanks{Booth School of Business, University of Chicago.  \it{kenichiro.mcalinn@chicagobooth.edu}}, Asahi Ushio\thanks{Faculty of Science and Engineering, Keio University. \it{ushioasahi@keio.jp}}, and Teruo Nakatsuma\thanks{Faculty of Economics, Keio University. \it{nakatuma@econ.keio.ac.jp}}
}

\maketitle\thispagestyle{empty}\setcounter{page}0

\begin{abstract}
\noindent 
The leverage effect-- the  correlation between an asset's return and its volatility-- has played a key role in forecasting and understanding volatility and risk.
While it is a long standing consensus that leverage effects exist and improve forecasts, empirical evidence paradoxically do not show that most individual stocks exhibit this phenomena, mischaracterizing risk and therefore leading to poor predictive performance.
We examine this paradox, with the goal to improve density forecasts, by relaxing the assumption of linearity in the leverage effect.
Nonlinear generalizations of the leverage effect are proposed within the Bayesian stochastic volatility framework in order to capture flexible leverage structures, where small fluctuations in prices have a different effect from large shocks.
Efficient Bayesian sequential computation is developed and implemented to estimate this effect in a practical, on-line manner.
Examining 615 stocks that comprise the S\&P500 and Nikkei 225, we find that relaxing the linear assumption to our proposed nonlinear leverage effect function improves predictive performances for 89\% of all stocks compared to the conventional model assumption.

\end{abstract}

\newpage
\section{Introduction \label{sec:intro} }
The estimation, inference, and prediction of volatility is one of the most crucial aspects in analyzing data with variability in order to make informed decisions.
In the field of finance and economics, volatility of financial assets has been investigated with great scrutiny to further the understanding of the mechanics and structure of price movement.
One aspect of volatility that has gathered special interest is the correlation between an asset's return and its volatility; coined the \emph{leverage effect}.
In particular, for decision making involving predictions, this correlation is critical, as knowing how today's change will effect tomorrow's risk is simply necessary for most sequential decision problems, especially under considerable shocks.
It is often claimed that this correlation is negative, implying that a negative (positive) shock to an asset's return results in an increase (decrease) in its volatility.
Thus, changing decisions accordingly based on predictions of increased or decreased volatility, implied by the previous shock.

This phenomenon is intuitive, as we can expect-- and often observe-- that an asset under distress exhibits more variability and uncertainty compared to an asset that is stable or increasing in price.
The term \emph{leverage} refers to an economic interpretation given by \citet{B76} and \citet{C82}. 
They state that, when an asset's price declines, the company's relative debt increases, making the balance sheet \emph{leveraged}, resulting in the company being riskier and therefore its market value more volatile (see \citealt{BW00}, for example, for different interpretations and comparisons of the leverage effect). 
Though only a hypothesis, this explanation has held weight in the field and the effect is widely believed to exist, with supporting evidence from examining major stock indices (\citealt{Nelson1991,Glosten1993,Dumas1998}, for ARCH-type models and \citealt{JPR94,WH97,JPR04,J05,OCSJ07,Nakajima2009,Nakajima2012,Takahashi2013,Shirota2014}, for SV-type models). 
However, contrary to consensus, the lack of empirical evidence of the effect from individual stocks is paradoxical; with most stocks exhibiting zero or very weak correlation between asset returns and volatility.
This is troublesome for decision makers wishing to exploit this structure, since mischaracterization of this correlation can lead to considerable loss in utility.

We postulate that this is caused by the simple, but almost universal, representation of the correlation: Most volatility models in the literature, basic or advanced, assume that the relationship between an asset's return and its volatility is linear, even though many advances have been made on other aspects of the model.
However, it is counter-intuitive to think that a large shock in return effects volatility with the same linear relationship as small daily fluctuations.
This notion has promoted research in considering more complex leverage effects.
For example, \citet{HHS11} introduced a more general form of the leverage effect by using a leverage function  within the GARCH framework.
In the context of stochastic volatility (SV) models, there has been no advances in this direction, even though SV models are known to outperform ARCH-type models due to its flexibility in capturing traits seen in asset returns \citep{Geweke1994,Fridman1998,Kim1998}.
The advances are hindered, partly, due to the computational complexity SV models entail, as it requires complex Markov chain Monte Carlo (MCMC) methods that are hard to sample and tune.

We respond to this movement by extending the SV model to include a leverage function in the form of a Hermite polynomial to examine the nonlinear dynamics of the correlation between an asset's return and its volatility.
To achieve this, we develop an effective Bayesian computation method using sequential Monte Carlo (SMC) by extending the particle learning method of \cite{CJLP10}, enabling estimation of the parameters of interest in a fast, efficient, and on-line manner that would have been previously near impossible.
With the new model and algorithm, we are able to examine and analyze the leverage effect over a large number of equity assets and over time, and find strong evidence for the leverage effect where it is unobserved, or weak, under the simple linear representation, with improved predictive performance.

We will define the SV model with nonlinear leverage functions and its estimation method in Section~\ref{sec:svm}.
Section~\ref{sec:study} will present the empirical study where we apply our model to daily returns of all stocks that compose the S\&P500 and Nikkei 225, with additional comments and further discussions in Section~\ref{sec:conc}.

\section{Stochastic Volatility Model with Leverage Functions \label{sec:svm} }
\subsection{Stochastic Volatility Non-Linear Leverage (SV-NL) \label{sec:svnl} }
The basic SV model is given by the following nonlinear dynamic model \citep{WH97,Carvalho2007,Lopes2010},
\begin{equation}
\label{sv}
\begin{cases}
 y_{t} = \exp\left(\frac{x_{t}}{2} \right)\epsilon_{t}, \\
 x_{t} = \mu + \beta x_{t-1} + \eta_{t},
\end{cases}
\begin{bmatrix} \epsilon_t \\ \eta_t \end{bmatrix}
\sim i.i.d.\:\mathcal{N}\left(
\begin{bmatrix} 0 \\ 0 \end{bmatrix},
\begin{bmatrix} 1 & 0 \\ 0 & \tau^2 \end{bmatrix}
\right),
\end{equation}
where $y_{t}$ is the asset return on day $t$ and $x_{t}$ is the time-varying log volatility.
The SV model is a state space model with observation noise $\epsilon_{t}$ and state noise $\eta_{t}$.
Both $\epsilon_{t}$ and $\eta_{t}$ are mutually and serially independent in this model represented by the off diagonal elements in the covariance matrix being zero. 
\citet{J05} compares two types of SV models with leverage (addressed as asymmetric SV in the paper) to represent the leverage effect in the model. 
The widely used of the two assumes the correlation between $\epsilon_{t}$ and $\eta_{t}$ as the following
\begin{equation}
\label{svl1}
\begin{cases}
 y_{t} = \exp\left(\frac{x_{t}}{2} \right)\epsilon_{t}, \\
 x_{t} = \mu + \beta x_{t-1} + \eta_{t},
\end{cases}
\begin{bmatrix} \epsilon_t \\ \eta_t \end{bmatrix}
\sim i.i.d.\:\mathcal{N}\left(
\begin{bmatrix} 0 \\ 0 \end{bmatrix},
\begin{bmatrix} 1 & \rho\tau \\ \rho\tau & \tau^2 \end{bmatrix}
\right),
\end{equation}
where $\rho$ is the correlation between the observation noise $\epsilon_{t}$ and the state noise $\eta_{t}$, allowing for a contemporaneous dependence in the variance. 

The state noise $\eta_{t}$ can then be rewritten as
\begin{eqnarray}
\eta_{t} = \rho\tau\epsilon_{t}+\sqrtsign{1-\rho^2}\tau u_{t}.
\end{eqnarray}
where $u_{t} \sim \mathcal {N}(0,1)$.
Hence, eqn.~\eqref{svl1} can be rewritten as a nonlinear state space model with uncorrelated observation noise $u_{t}$ and state noise $\epsilon_{t}$: 
\begin{equation}
\label{svl2}
\begin{cases}
 y_{t} = \exp\left(\frac{x_{t}}{2} \right)\epsilon_{t}, \\
 x_{t} = \mu + \beta x_{t-1} + \varphi \epsilon_{t-1} + \omega u_{t},
\end{cases}
\begin{bmatrix} \epsilon_t \\ u_t \end{bmatrix}
\sim i.i.d.\:\mathcal{N}\left(
\begin{bmatrix} 0 \\ 0 \end{bmatrix},
\begin{bmatrix} 1 & 0 \\ 0 & 1 \end{bmatrix}
\right),
\end{equation}
where $\varphi = \rho\tau$, and $\omega = \sqrtsign{1-\rho ^2}\tau$.

The model representation of eqn.~\eqref{svl2} assumes linear correlation, as in the term, $\varphi \epsilon_{t-1}$, is linear, though the assumption is not founded in evidence nor theory but in convenience. 
We extend the model to include a nonlinear leverage function, $\ell(\cdot)$: 
\begin{equation}
\label{svl.nonlinear}
\begin{cases}
 y_{t} = \exp\left(\frac{x_{t}}{2} \right)\epsilon_{t}, \\
 x_{t} = \mu + \beta x_{t-1} +  \ell(\epsilon_{t-1}) + \omega u_{t},
\end{cases}
\begin{bmatrix} \epsilon_t \\ u_t \end{bmatrix}
\sim i.i.d.\:\mathcal{N}\left(
\begin{bmatrix} 0 \\ 0 \end{bmatrix},
\begin{bmatrix} 1 & 0 \\ 0 & 1 \end{bmatrix}
\right),
\end{equation}
where the nonlinear leverage function $\ell(\cdot)$ can be any function that transmits an impact of the previous shock in the asset price $\epsilon_{t-1}$ onto the log volatility $x_t$. 
If $\ell(\epsilon_{t-1})$ is a linear function, say $\varphi\epsilon_{t-1}$, the SV model with nonlinear leverage in eqn.~\eqref{svl.nonlinear} is reduced to eqn.~\eqref{svl2}. 
Although the exact functional form of $\ell(\cdot)$ is unclear, we do not want to impose parametric assumptions upon it. 
We instead introduce a more flexible polynomial approximation of the leverage function. 

The $k$-th order Hermite polynomial $H_{k}(z)$ is defined as
\begin{equation}
\label{hermite}
H_{k}(z)=(-1)^k\exp\left(\frac{z^2}{2}\right)\frac{d^k}{dz^k}\exp\left(-\frac{z^2}{2}\right),
\end{equation}
where the first seven Hermite polynomials are
\begin{align*}
H_{0}(z) &= 1, \quad
H_{1}(z) = z, \quad
H_{2}(z) = z^2-1, \quad
H_{3}(z) = z^3-3z, \\ 
H_{4}(z) &= z^4-6z^2, \quad 
H_{5}(z) = z^5-10z^3+15z, \quad 
H_{6}(z) = z^6-15z^4+45z^2-15.
\end{align*}

Hermite polynomials have some desirable properties for the analysis of leverage effects:
\begin{enumerate}
\item
Zero Expectation:\\
When $z$ follows a normal distribution with zero expected value and unit variance, the expected value of the Hermite polynomial is equal to zero for any $k$. In other words,
\[
E(H_{k}(z))=\int_{-\infty}^{\infty} H_{k}(z)\phi(z) dz=0,\:\:\:
z \sim N(0,1),
\]
where $\phi(z) \propto \exp(-\frac{z^2}{2})$.

\item
Orthogonality:\\
Hermite polynomials are orthogonal to each other with respect to the weight function $\phi(z) \propto \exp(-\frac{z^2}{2})$. Thus,
\[
E(H_{j}(z) H_{k}(z))= \int_{-\infty}^\infty H_j(z)H_k(z)\phi(z)dz
 = 
 \begin{cases}
 n! & (k=j); \\
 0 & (k\ne j).
 \end{cases}
\]
\end{enumerate}
Given the two properties above, they form the orthogonal basis of the Hilbert space of leverage functions such that $\int_{-\infty}^\infty|\ell(z)|^2\phi(z)dz<\infty$. 

We now define the approximated leverage function using Hermite polynomials as
\begin{equation}
\label{svnlv.hermite.nl}
 \ell^H_k(\epsilon_{t-1}) := 
\varphi_1 H_1(\epsilon_{t-1}) + \dots + \varphi_k H_k(\epsilon_{t-1}). 
\end{equation}

Finally, the SV model with the $k$-th order nonlinear leverage function is defined as
\begin{equation}
\label{svl3}
\begin{cases}
 y_{t} = \exp\left(\frac{x_{t}}{2} \right)\epsilon_{t}, \\
 x_{t} = \mu + \beta x_{t-1} +  \ell^H_k(\epsilon_{t-1}) + \omega u_{t},
\end{cases}
\begin{bmatrix} \epsilon_t \\ u_t \end{bmatrix}
\sim i.i.d.\:\mathcal{N}\left(
\begin{bmatrix} 0 \\ 0 \end{bmatrix},
\begin{bmatrix} 1 & 0 \\ 0 & 1 \end{bmatrix}
\right).
\end{equation}
This SV model with nonlinear leverage, or SV-NL model in short, is the model that will be used in our empirical study of leverage effects in individual stock price returns (Section~\ref{sec:study}).

\subsection{Bayesian Sequential Computation \label{sec:PL}}
In this section, we briefly elaborate our proposed particle learning method for the SV-NL model in eqn.~\eqref{svl3}. 
In general, there are two methods for Bayesian computation of SV models: MCMC based \citep{Chib2002,OCSJ07} and SMC based \citep{Polson2004,Lopes2011,Creal2012}.
While both methods have their pros and cons (see, for example, \citealt{Lopes2006}), there is merit in using SMC in financial applications because of its on-line nature (i.e. the ability to produce forecasts fast), which is appealing to finance practitioners.
However, SMC methods struggled with the problem of parameter learning \citep{GSS93,K96,PS99}  and multiple methods have been proposed \citep{LW01,Storvik02,Fearnhead02,PSM08,JP08,JPY08}.
We will not review the literature of SMC methods, instead we will expand on the most recent development by \cite{CJLP10}, which enables parameter leaning in nonlinear, non-Gaussian models under certain conditions.

Parameters of interest in the model are unknown in most applications of finance and econometrics, and for this particular model in eqn.~\eqref{svl3}, inferring on the parameters in the leverage function is of the utmost importance. 
When a state space model depends on unknown but static parameters $\theta$, we have the following state space representation,
\begin{eqnarray}
\left\{
\begin{array}{l}
y_t \sim p(y_t \vert x_t,\theta), \\
x_t \sim p(x_t \vert x_{t-1},\theta),
\end{array}
\right.
\end{eqnarray}
where we need to evaluate the posterior distribution of $\theta$ given the observations $y_{1:t}$, $p(\theta|y_{1:t})$, as well as the latent state, $x_{1:T}$.
In the framework of SMC methods, $p(\theta|y_{1:t})$ is sequentially updated as new observations arrive in an on-line manner. 
This is called \emph{particle learning}, as we learn the parameters using particle filtering. 
A na{\"i}ve particle learning algorithm, which is a particle filter for the extended state vector $z_{t}=(x_{t},\theta)$, is defined as follows. 
Let $\{\hat{z}_{t}^{(i)}=(\hat{x}_{t}^{(i)},\hat{\theta}_{t}^{(i)})\}_{i=1}^N$ and $\{\tilde z_t^{(i)}=(\tilde x_t^{(i)},\tilde \theta_t^{(i)})\}_{i=1}^N$ denote particles, $i=1:N$, jointly generated from $p(z_{t}|y_{1:t-1})$ and $p(z_{t}|y_{1:t})$, respectively. 
Then, the particle approximation of the Bayesian learning process is given by
\begin{align}
p(z_{t}|y_{1:t-1}) &\simeq \frac1N\sum_{i=1}^N p(z_{t}|\tilde z_{t-1}^{(i)}),\\
p(z_t\vert y_{1:t}) 
&\simeq \sum_{i=1}^{N}W^{(i)}_{t}\delta(z_{t}-\hat{z}_{t}^{(i)}), \quad
W^{(i)}_{t} = \displaystyle 
\frac{p(y_{t}\vert \hat{z}_{t}^{(i)})} {\sum^{N}_{i=1}p(y_{t}\vert \hat{z}_t^{(i)})}.
\end{align}

Although the above learning algorithm is easy to implement, researchers have been aware that, with the na{\"i}ve particle learning algorithm, particles tend to degenerate in the process of resampling. 
As a more efficient alternative, \citet{CJLP10} propose a particle learning (PL) algorithm that resamples and updates the sufficient statistics of the posterior distributions of the parameters. 

However, in order to apply the particle learning algorithm, we need to know the functional form of $p(y_{t}|x_{t-1},\theta)$ and be able to draw $x_{t}$ from $p(x_{t}|x_{t-1},y_{t},\theta)$. 
For the SV-NL model in eqn.~\eqref{svl3}, however, neither is possible. 
To circumvent these problems, we adopt the \emph{auxiliary particle filter} by \citet{PS99} within the particle learning framework.

According to \citet{PS99}, SMC methods provide good estimations of state variables and parameters, where the model is a good approximation of the data, and the conditional density $p(y_{t}|x_{t})$ is reasonably flat in $x_{t}$. 
They point out two weaknesses of the algorithm. 
Firstly, when there are outliers, SMC methods cannot precisely adapt, so that the state variables would be underestimated even when $N$ is large. 
In this case, the variability of the likelihood, i.e. $p(y_{t}|x_{t})$, would be increased, which reduces the precision of resampling because it is based on the weight proportional to $p(y_{t}|x_{t})$. 
Secondly, degeneration of particles are a general problem that occurs when the likelihood is similar in each cycle of the particle filter. 
It causes poor tail representation of the predictive density because the particles placed on the tails are assigned with similarly low weights, increasingly reducing the weights to zero as the algorithm furthers. 
As a result, the particles may degenerate to a few points. 
The auxiliary particle filter softens the influence of outliers because the second-stage resampling would be much less variable than the original sampling. 
Moreover, the first-stage resampling is based on the current observation value $y_{t}$ so we can expect that the good particles are likely to be propagated forward. 

Finally, the algorithm for particle learning with auxiliary variables (PLAV) is summarized as follows.
\newline
\newline
\textsc{Algorithm: Particle Learning with Auxiliary Variables}
\vspace{-0.2cm}
\begin{description}
\item[\hspace{0.75cm}Step 0:] Sample the starting values of $N$ particles $\{\tilde{z}_{0}^{(i)}\}^{N}_{i=1}$ from $p(z_0)$.
\vspace{-0.2cm}
\item[\hspace{0.75cm}Step 1:] Resample $\{\hat{z}^{(i)}_{t-1}\}^{N}_{i=1}$ from $\{\tilde{z}^{(i)}_{t-1}\}_{i=1}^{N}$ with $\hat{W}_{t}^{(i)}\propto p(y_{t}|g(\tilde{x}_{t-1}^{(i)},\tilde{\theta}_{t-1}^{(i)}))$.
\vspace{-0.2cm}
\item[\hspace{0.75cm}Step 2:] Propagate $\hat{x}^{(i)}_{t}$ from $p(x_{t}|\hat{x}^{(i)}_{t-1},\hat{\theta}_{t-1}^{(i)})$.
\vspace{-0.2cm}
\item[\hspace{0.75cm}Step 3:] Resample $\{\tilde{x}^{(i)}_{t}\}^{N}_{i=1}$ from $\{\hat{x}^{(i)}_{t}\}^{N}_{i=1}$ with $W_{t}^{(i)}\propto \displaystyle\frac{p(y_{t}|\hat{x}^{(i)}_t)}{p(y_{t}|g(\hat{x}^{(i)}_{t-1},\hat{\theta}_{t-1}^{(i)}))}$.
\vspace{-0.2cm}
\item[\hspace{0.75cm}Step 4:] Update the sufficient statistics $\tilde{s}^{(i)}_{t}=\mathcal{S}(\tilde{s}^{(i)}_{t-1},\hat{x}^{(i)}_{t},y_{t})$, $i\in\{1,\dots,N\}$.
\vspace{-0.2cm}
\item[\hspace{0.75cm}Step 5:] Sample $\tilde{\theta}_{t}^{(i)}$ from $p(\theta|\tilde{s}^{(i)}_{t})$, $i\in\{1,\dots,N\}$.
\end{description} 
\hspace*{\fill} 


\subsection{Particle learning with auxiliary variables: A simulation example}

Before we test the algorithm on actual data, we will compare the results from PLAV with PL and an MCMC alternative \citep{OCSJ07}.
Since PL cannot estimate the parameters in the linear-leverage SV model in eqn.~\eqref{svl2}, we will compare the no-leverage SV model (eqn.~\ref{sv}) between the three and linear-leverage SV model (eqn.~\ref{svl2}) between PLAV and MCMC.
Following \citet{OCSJ07}, we will generate data from the linear-leverage SV model in eqn.~\eqref{svl2} with parameters $\mu= -0.026,\:\:\beta= 0.970,\:\:\varphi= -0.045,\:\:\omega=  0.143$ and the no-leverage SV model with parameters $\mu= -0.026,\:\:\beta= 0.970,\:\:\tau = 0.150$. 
The posterior mean and the credible interval of each parameter for the no-leverage SV model and the linear-leverage SV model are given in Tables~\ref{pmsv} and \ref{pmlsv}, respectively. 
We can see that the estimation results of both SV models, with and without leverage, by three algorithms are comparable and their credible intervals mostly cover the assigned true values (the exception being the MCMC estimation of $\mu$ for the linear-leverage MCMC model).
 The execution time for each algorithm is given in Table~\ref{timesv}. 
 The results in Table~\ref{timesv} show that MCMC takes much less time for the no-leverage SV model compared to PL and PLAV. 
 However, for the linear-leverage SV model, MCMC takes about nine times as much as PLAV.
 From this, we can conclude that PLAV is preferred over MCMC in terms of on-line estimation often seen in practice and especially if the model estimated is more complex than the no-leverage SV model.

\section{Empirical Study \label{sec:study}}
\subsection{Data}
The object of our empirical study is to forecast the volatility of stock returns  using the SV-NL model of eqn.~\eqref{svl3} for individual stocks that comprise the two major stock indices: The S\&P 500 and Nikkei 225.
 Daily closing prices of each stock from the beginning of 2004 to the end of 2013 are used for the analysis, with the first two years used as a training period, and the rest analyzed sequentially over the eight year period.
 This is done to mirror the real life decision making process encountered with this type of data.
 Since some of the stocks were not listed for the entire time period, they were excluded. 
 A total of 615 stocks were analyzed with 417 stocks from the S\&P 500 and 198 stocks from the Nikkei 225.
 By doing a large scale analysis of this magnitude, we are not only able to explore and compare leverage effects and its predictive performance, but also check for robustness of the SV-NL across multiple markets and industries. 
 
\subsection{Model comparison and prior specifications}

For each stock return series, we estimate the SV-NL model in eqn.~\eqref{svl3} from the 0th-order (no-leverage) up to the 6th-order leverage function. 
In order to discern the best order for prediction, we compare the one-step ahead predictive marginal likelihood:
\begin{equation}
p(y_{1:T}|k)=\prod_{t=1}^{T}p(y_{t}|y_{1:t-1},k)=\prod_{t=1}^{T}\int p(y_t|x_t)p(x_t|y_{1:t-1},k)dx_t,\quad k\in\{0,1,\dots,6\}, 
\end{equation}
for each $t=1:T$ and order $k$; the order of the Hermite polynomial. 
In the Bayesian model selection procedure, the model with the highest predictive marginal likelihood at time $t$ is selected as the best model representing the data up to time $t$.
Thus, at time $T$, we are choosing the best model for forecasting through the whole period of examination.
In the framework of particle learning, this can be calculated directly during the filtering procedure with
\begin{equation}
p(y_{1:T}|k) \simeq \prod_{t=1}^{T}\left\{\frac1{N}\sum_{i=1}^{N}p(y_{t}|x_{t,k}^{(i)})\right\},
\end{equation}
where $x_{t,k}^{(i)}$ $(i\in\{1,\dots,N\})$ is the particle $i$ generated from $p(x_{t}|\tilde{x}_{t-1}^{(i)},\tilde{\theta}_{t-1}^{(i)},k)$, which is the density of the state equation in the SV-NL model in eqn.~\eqref{svl3}. 
We set $N=10,000$ and evaluate the marginal likelihood with the last eight years of the ten years of data, using the first two years as the learning period. 
In executing the particle learning algorithm, we set the prior specifications as
$A_{0} = {\rm diag}\bigl\{ 1.0,0.01,1.0,\cdots,1.0\bigl\}, b_{0}^{'} = [0.0,\:0.95,\:0.0,\cdots 0.0],c_{0} = 5,\:\:d_{0} = 0.4,$ which are relatively non-informative priors standard in the literature.

\subsection{Empirical results\label{sec:emp}}

\begin{figure}[hbt]
\begin{center}
\includegraphics[width=12cm,clip]{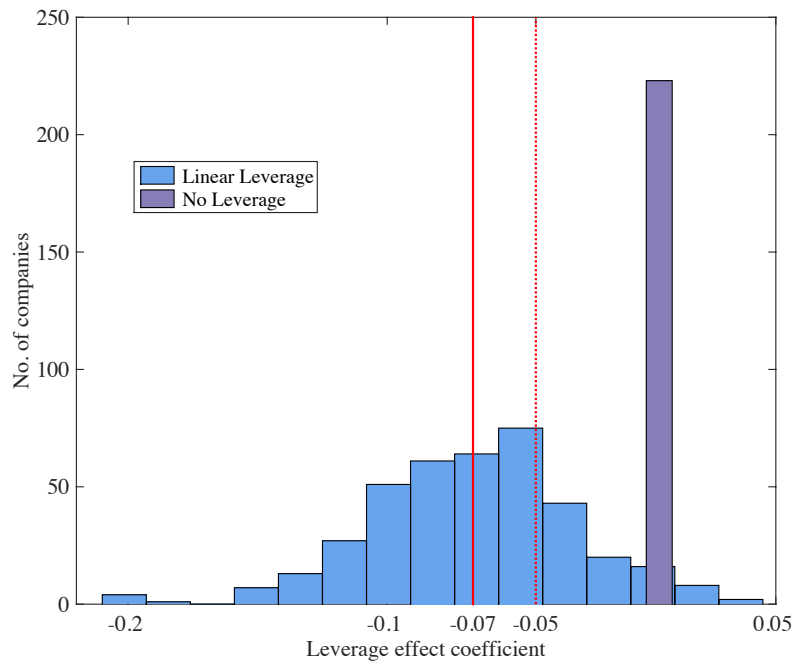}
\caption{Histogram of leverage effect coefficients under the linear leverage effect assumption. Stocks that are chosen to have leverage effects are in light blue and stocks that are chosen not to have any leverage effects are in purple. Red lines denote the average leverage effect excluding (solid) and including (dotted) stocks with no leverage.}
\label{leverage}
\end{center}
\end{figure}

\begin{figure}[t]
\begin{center}
\includegraphics[width=10cm,clip]{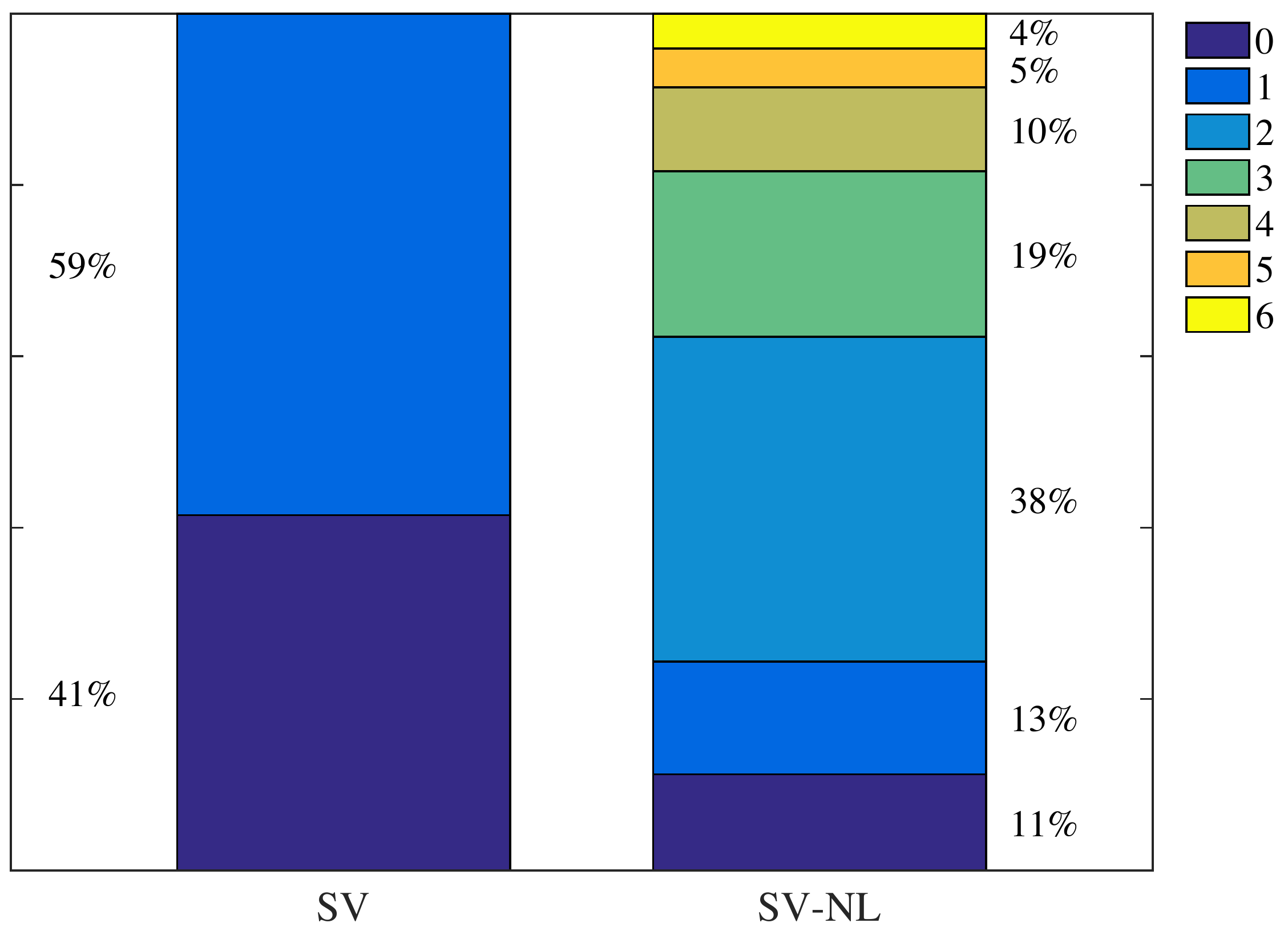}
\caption{Optimal leverage orders selected using the predictive marginal likelihood under the assumption of linear leverage (left) and nonlinear leverage (right)}
\label{ratio}
\end{center}
\end{figure}

\paragraph{Comparing against the conventional assumption.} We will first examine the predictive performance of individual stocks under the linear assumption to see whether the leverage effect contributes to its predictive performance.
To do this, we limit the leverage order to zero and one and examine whether each stock chooses models in eqn.~\eqref{sv} or eqn.~\eqref{svl2} under the same specifications.
Examining the 615 stocks, we find that the no-leverage model (eqn.~\ref{sv}) is chosen as the best predictive model for nearly half of the stocks and the other half showing weak leverage effects (Figure.~\ref{leverage}, Figure.~\ref{ratio}: left column).
Compared to the results reported in other papers (-0.3179 in \citealt{J05}, -0.3617 in \citealt{OCSJ07}, -0.4825 in \citealt{Nakajima2009}), we find that the average leverage effect to be much smaller (-0.07 if we exclude the zeros and -0.05 if we include the zeros). 
This confirms that, under the standard linear assumption, most stocks either appear not to exhibit a leverage effect or are very small.
This might lead the decision maker to then assume that the leverage effect does not exist, or worthwhile modeling due to its increased computational complexity.
However, this assumption, as we will now show, is incorrect and potentially damaging to the forecasts and subsequent decisions being considered.

Relaxing the models to include higher order leverage (Figure.~\ref{ratio}: right column), we find that, in terms of predictive performance, the no-leverage is chosen as the best model for only 11\% under the generalized framework, meaning that around 30\% were misspecified, as well as 46\% being misspecified as linear leverage, making the total of misspecified models to be around 76\%, that is 76\% were losing predictive performance under this setup.
This reversal strongly indicates that the conventional linear assumption is misleading and biases estimation results towards evidence of no leverage effect, and thus leading to poorer predictive performances.

Given this result, we can see that the leverage effect is much more persistent in predicting future volatility than previously expected. 
For the stocks analyzed, models with some order of leverage effect (89\%) shows improvement in predictive performance over no leverage and we can see that many of them have higher order leverage (76\%).

\paragraph{Predictive performance.}
We further our analysis by examining the amount of improvements made by relaxing the linear assumption in the leverage effect.
Here, we compare the difference in log predictive marginal likelihood between the best model chosen under the linear assumption and the best model chosen under the nonlinear extension.
This, in the literature, is called the log predictive density ratio,
\begin{align*}
	\mathrm{LPDR}_{1{:}t}(t+1)=\sum_{i=1{:}t}\mathrm{log}\{p_{nl}(y_{t+1}|y_{1{:}t})/p_{\mathrm{l}}(y_{t+1}|y_{1{:}t})\}
\end{align*}
where $p_{nl}(y_{t+1}|y_{1{:}t})$ is the predictive density under the SV model with nonlinear leverage and $p_{l}(y_{t+1}|y_{1{:}t})$ is the predictive density under the SV model with linear leverage.
As used by several authors recently~\cite[e.g.][]{Nakajima2010,Aastveit2015,McAlinn2016},  LPDR measures provide a direct
statistical assessment of relative distribution accuracy.
In this case, a positive LPDR indicates that the SV-NL outperforms SV under the linear assumption.

\begin{figure}[hbt]
\begin{center}
\includegraphics[width=12cm,clip]{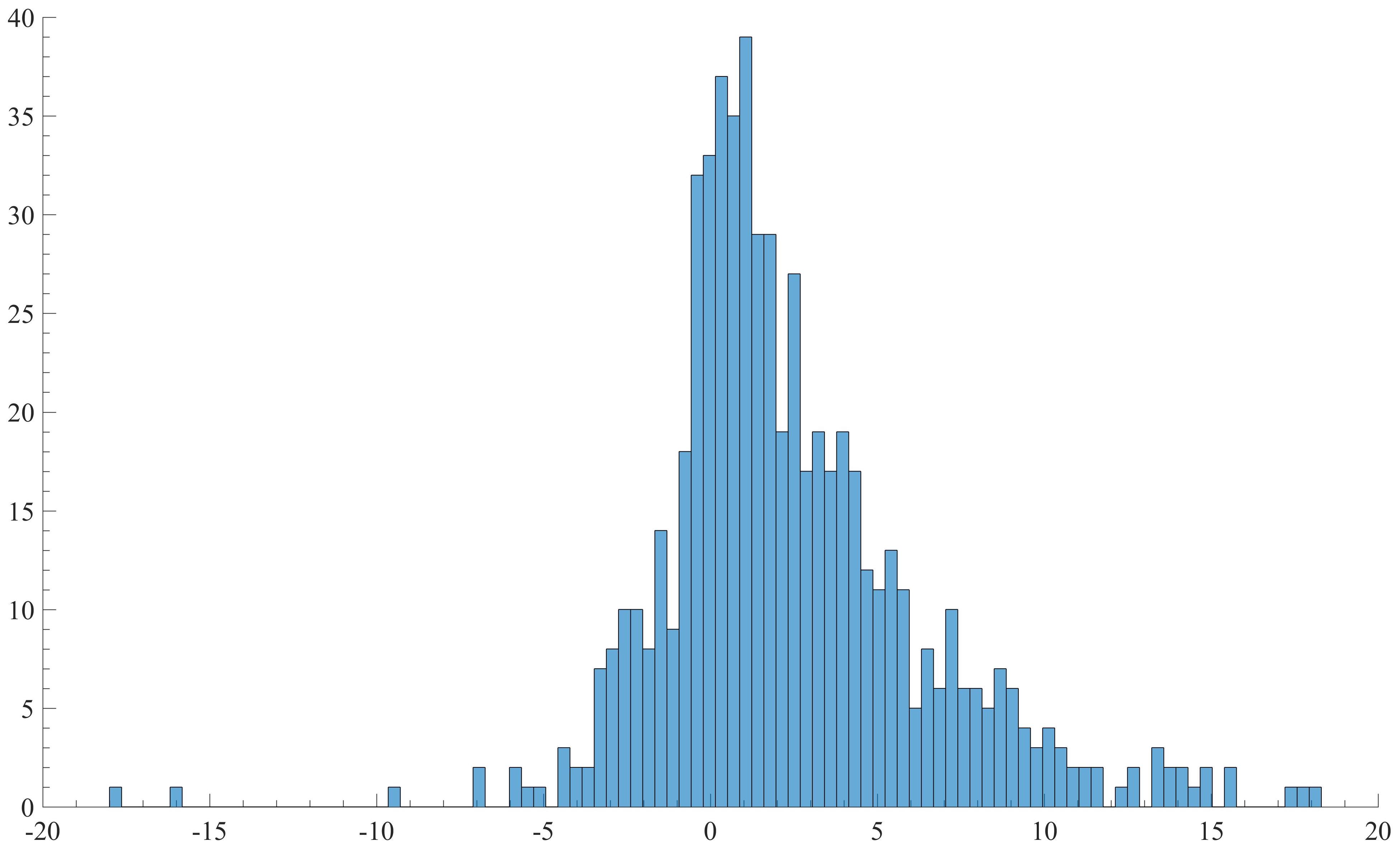}
\caption{Histogram of the log predictive marginal likelihood for all stocks compared between the best model within the linear leverage assumption (leverage order 0-1) and within the nonlinear leverage framework (leverage order 2-6). LPDR above zero indicates improvements over the linear leverage assumption.}
\label{mlALL}
\end{center}
\end{figure}

\begin{figure}[hbt]
\begin{center}
\includegraphics[width=12cm,clip]{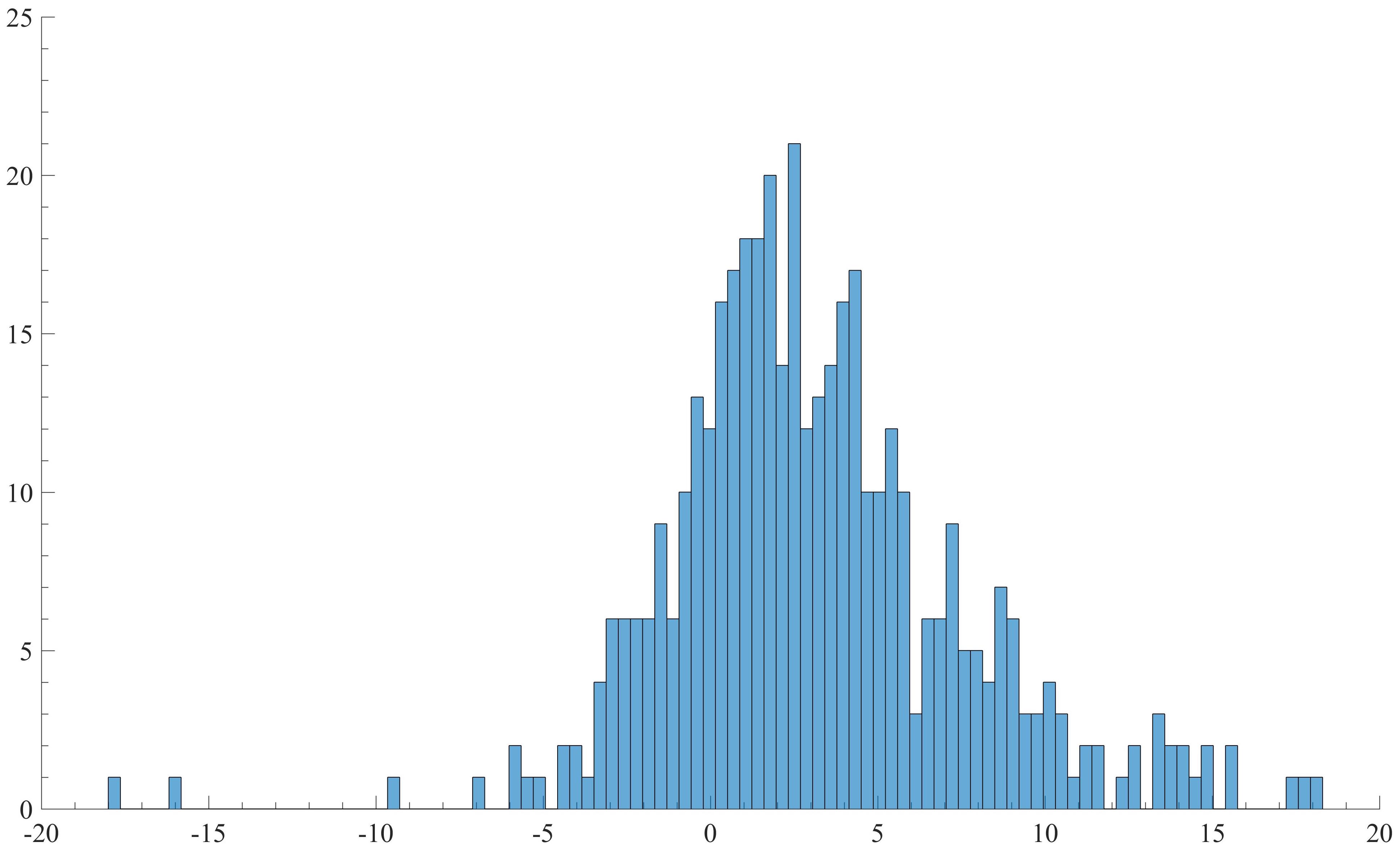}
\caption{Histogram of the log predictive marginal likelihood for stocks comprising the S\%P 500 compared between the best model within the linear leverage assumption (leverage order 0-1) and within the nonlinear leverage framework (leverage order 2-6). LPDR above zero indicates improvements over the linear leverage assumption.}
\label{mlUS}
\end{center}
\end{figure}

\begin{figure}[hbt]
\begin{center}
\includegraphics[width=12cm,clip]{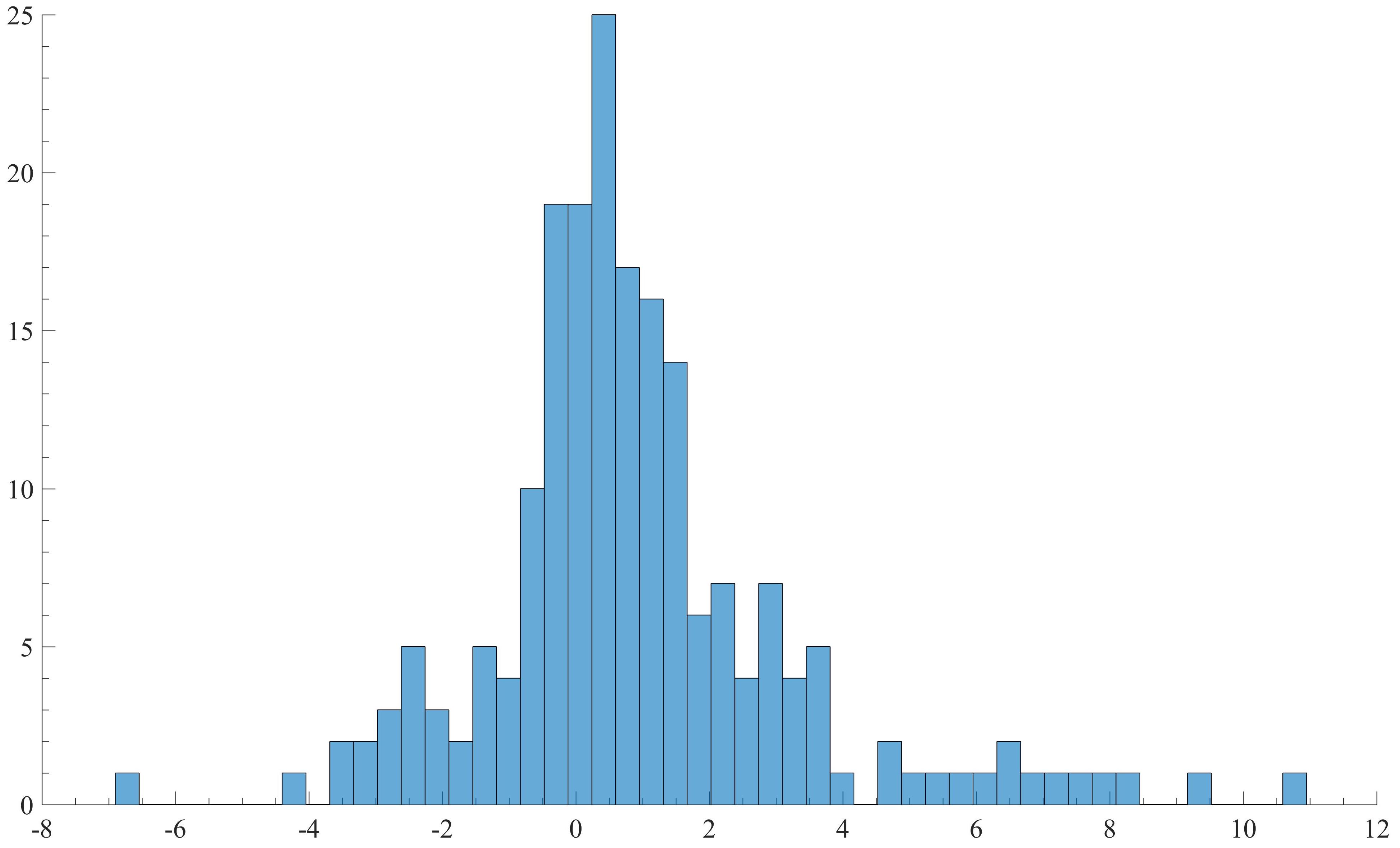}
\caption{Histogram of the log predictive marginal likelihood for stocks comprising the Nikkei 225 compared between the best model within the linear leverage assumption (leverage order 0-1) and within the nonlinear leverage framework (leverage order 2-6). LPDR above zero indicates improvements over the linear leverage assumption.}
\label{mlJP}
\end{center}
\end{figure}

We compare this for all stocks (Figure.~\ref{mlALL}), U.S. stocks (Figure.~\ref{mlUS}), and Japanese stocks (Figure.~\ref{mlJP}).
Note that, for the histograms only, we excluded Fidelity National Information Services Inc. and TOTO from their respective markets due to it heavily favoring zero leverage and skewing the proportion of histograms, though they do not effect the overall conclusions reached.
From the overall results with all assets, it is clear the predictive gains coming from using nonlinear leverage is significant, with a completely asymmetric distribution of LPDRs.
Additionally, while LPDR$<-10$ are rare, there are many more stocks with LPDR$>10$, making the asymmetry more profound.

Moving to the U.S. stocks, we find this asymmetry (or skewness) to be more evident, with a shift of the mode towards the positive (i.e. SV-NL performing better).
In contrast, Japanese stocks seem to be centered much more around zero, though it is still clear that there are significant gains for many of the stocks.
The difference between U.S. stocks and Japanese stocks are noteworthy, as it is obvious that U.S. stocks are more likely to gain from nonlinear leverage effects, and that the gains are larger.
This shows that, while the improvements in predictive performance is persistent across all stocks, certain markets are more likely to gain from the nonlinear framework.

\begin{figure}[t]
\begin{center}
\begin{tabular}{c}
\begin{minipage}{0.5\hsize}
\begin{center}
\includegraphics[width=7cm,clip]{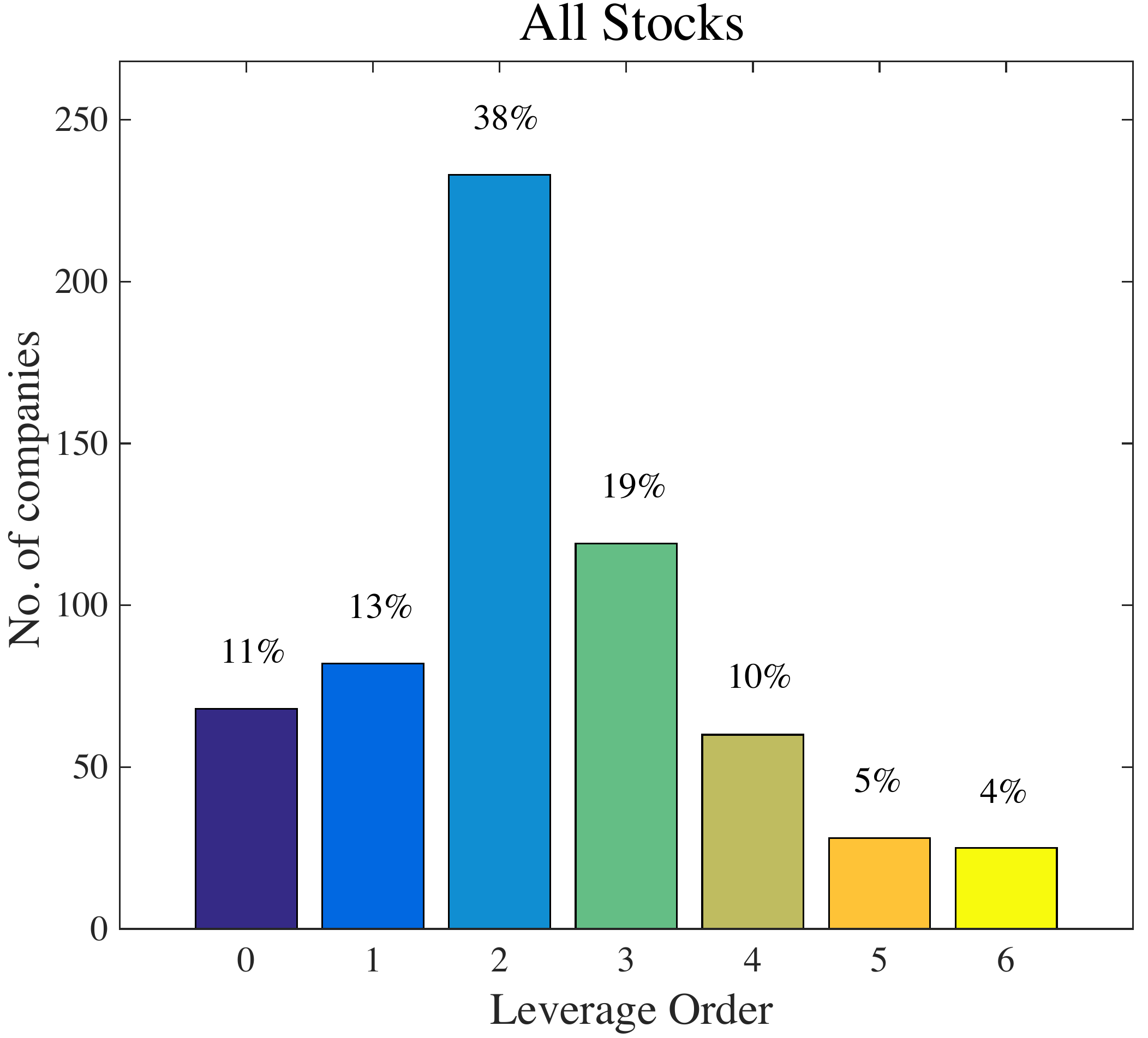}
\end{center}
\end{minipage}
\begin{minipage}{0.5\hsize}
\begin{center}
\includegraphics[width=7cm,clip]{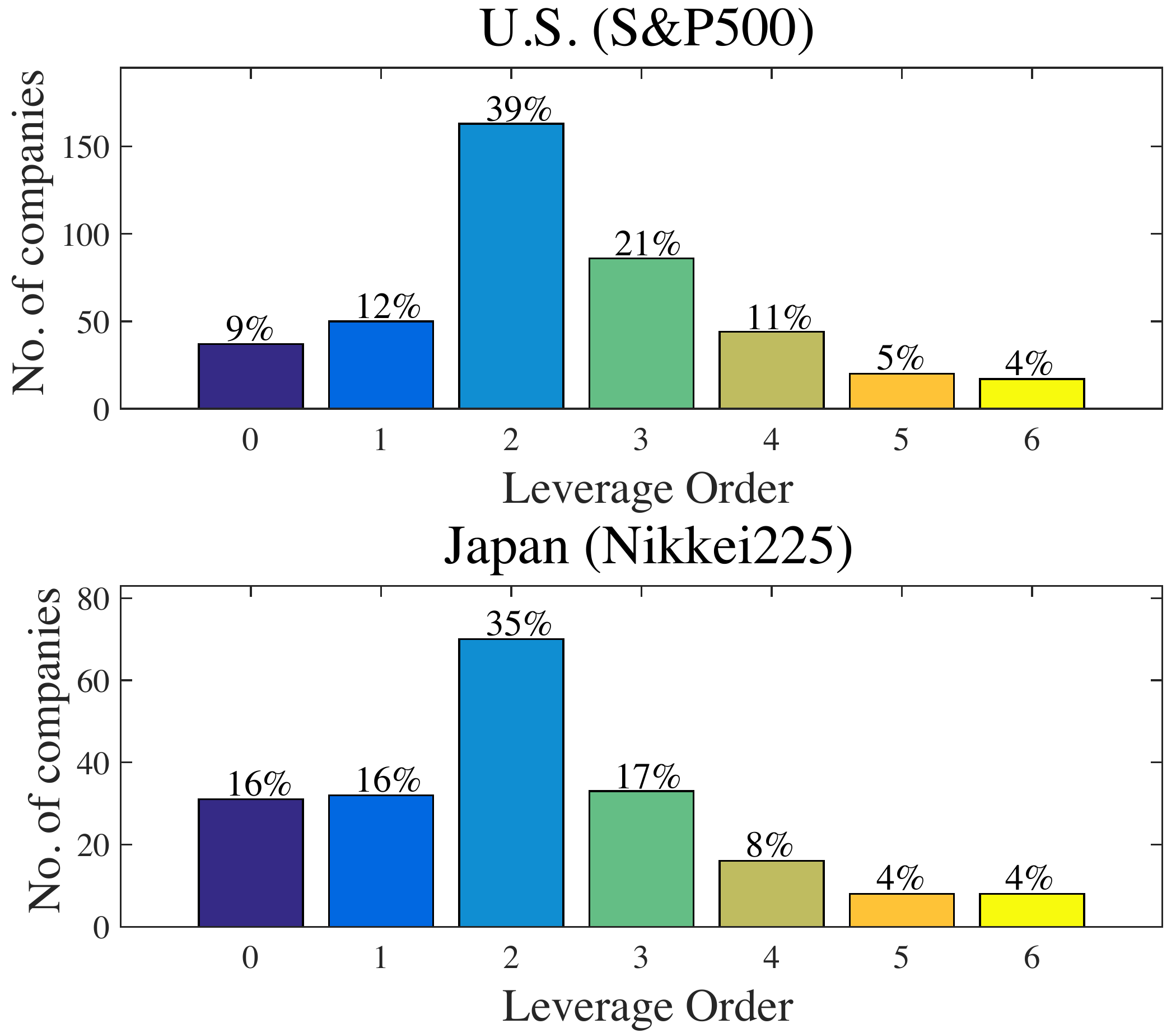}
\end{center}
\end{minipage}
\end{tabular}
\caption{Number of optimal orders of leverage functions selected using the predictive marginal likelihood for all stocks (left), S\&P 500 (top right), and Nikkei225 (bottom right).}
\label{all}
\end{center}
\end{figure}

\paragraph{Leverage characteristics.} 
We further our analysis by examining the characteristics of the leverage function in SV-NL.
Figure.~\ref{all} graphs the number of polynomial orders that are selected within a group using the marginal likelihood. 
The left panel shows the results for all 615 stocks, while the right panels show the results by market. 
As seen in the previous section, for all the stocks, we can see that many stocks improve predictive performance with 2nd-order leverage (38\%) followed by 3rd-order leverage (19\%).
In terms of higher order leverage (leverage in the order of 2-6), they improve for approximately 76\% of all stocks.
Leverage in the order of 4-6 improve for very little of the stocks examined, implying that the leverage is captured fairly simply by lower order functions and is not overfitting.

Similar patterns are observed in the right panels of Figure.~\ref{all}, which indicate that the nonlinear leverage effect improves performance and is persistent over both the U.S. and Japanese stocks, which indicate that it is not a market dependent phenomenon. 
However, even though the rough shapes of both histograms are similar, we can see that there is a slight difference between the two markets. 
For example, the percentage of stocks with no leverage or linear leverage performing the best is larger for stocks in the Nikkei 225 compared to the stocks in the S\&P 500. 
It is commonly perceived that Japanese companies and investors are more conservative than the U.S., and we conjecture that this difference in risk preference/aversion between the two countries' investors may play a role, to some degree, in the volatility structure of the stocks.
Nonetheless, higher order leverage functions are supported for the majority of the stocks in both markets, strongly suggesting that the effect exists and that it improves predictive performance, but not in a way (i.e. linear) we previously perceived.

\begin{figure}[t]
\begin{center}
\includegraphics[width=15cm,clip]{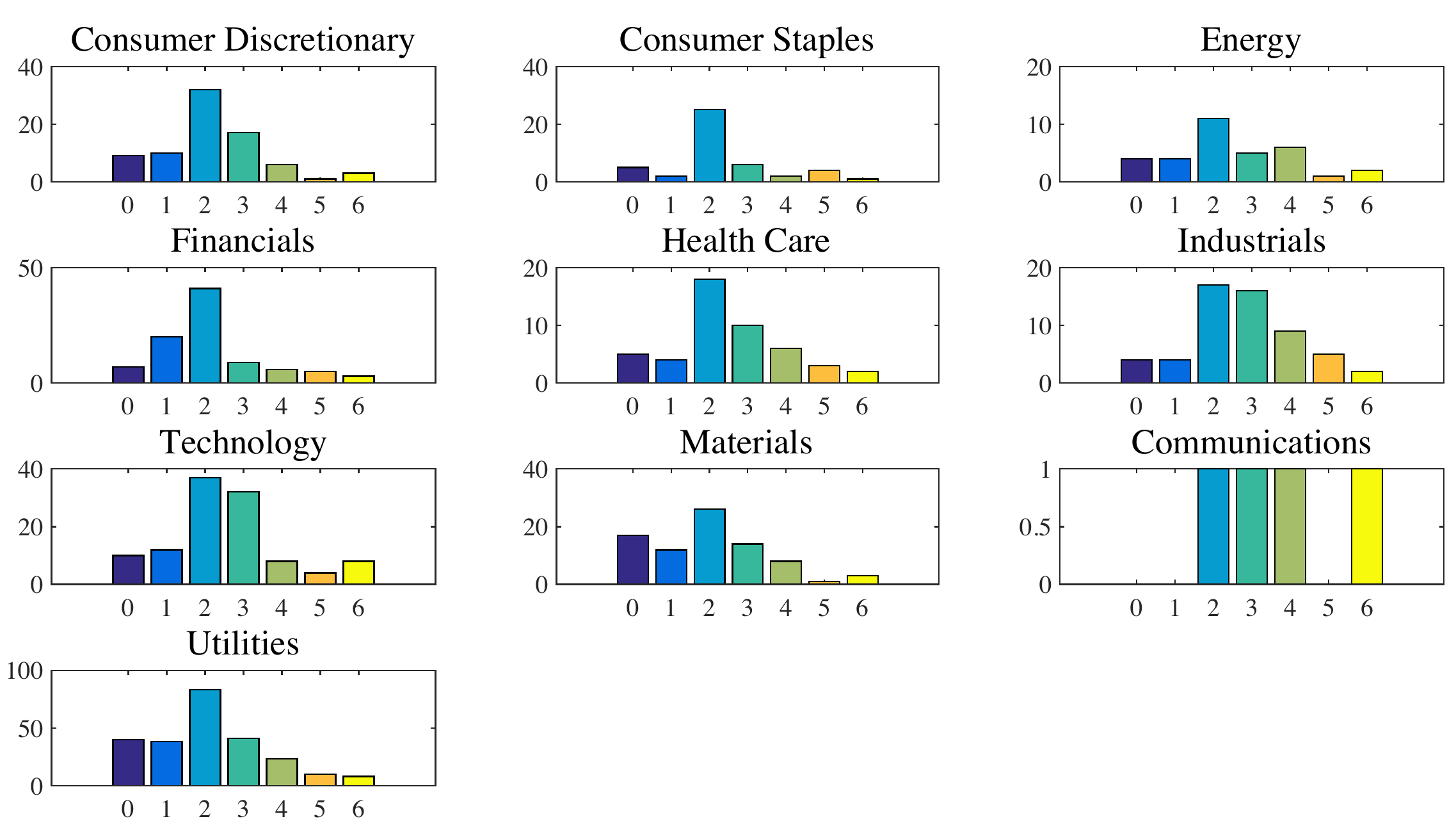}
\caption{Number of optimal orders of leverage functions selected using the predictive marginal likelihood for sectors.}
\label{fig7}
\end{center} 
\end{figure}

\paragraph{Sector results.} Looking closer at leverage order within sectors (Figure.~\ref{fig7}), we observe that, in all sectors, the 2nd-order leverage function is the most selected order-- consistent with the overall results-- except for \emph{Communications}, which only has four stocks (all higher order). 
Though similar, closer examination of the distribution of orders within sector suggests each sector to have some unique characteristics.
For example, sectors such as \emph{Health Care}, \emph{Industrials}, and \emph{Technology} have more high order leverage than low order (zero and one) leverage stocks, while \emph{Financials}, \emph{Materials}, and \emph{Utilities} having more low order leverage stocks.
These characteristics might be explained by such factors like industry size, investor perceptions, accounting rules, and so on.

\begin{figure}[t]
\begin{center}
\includegraphics[width=5.6in]{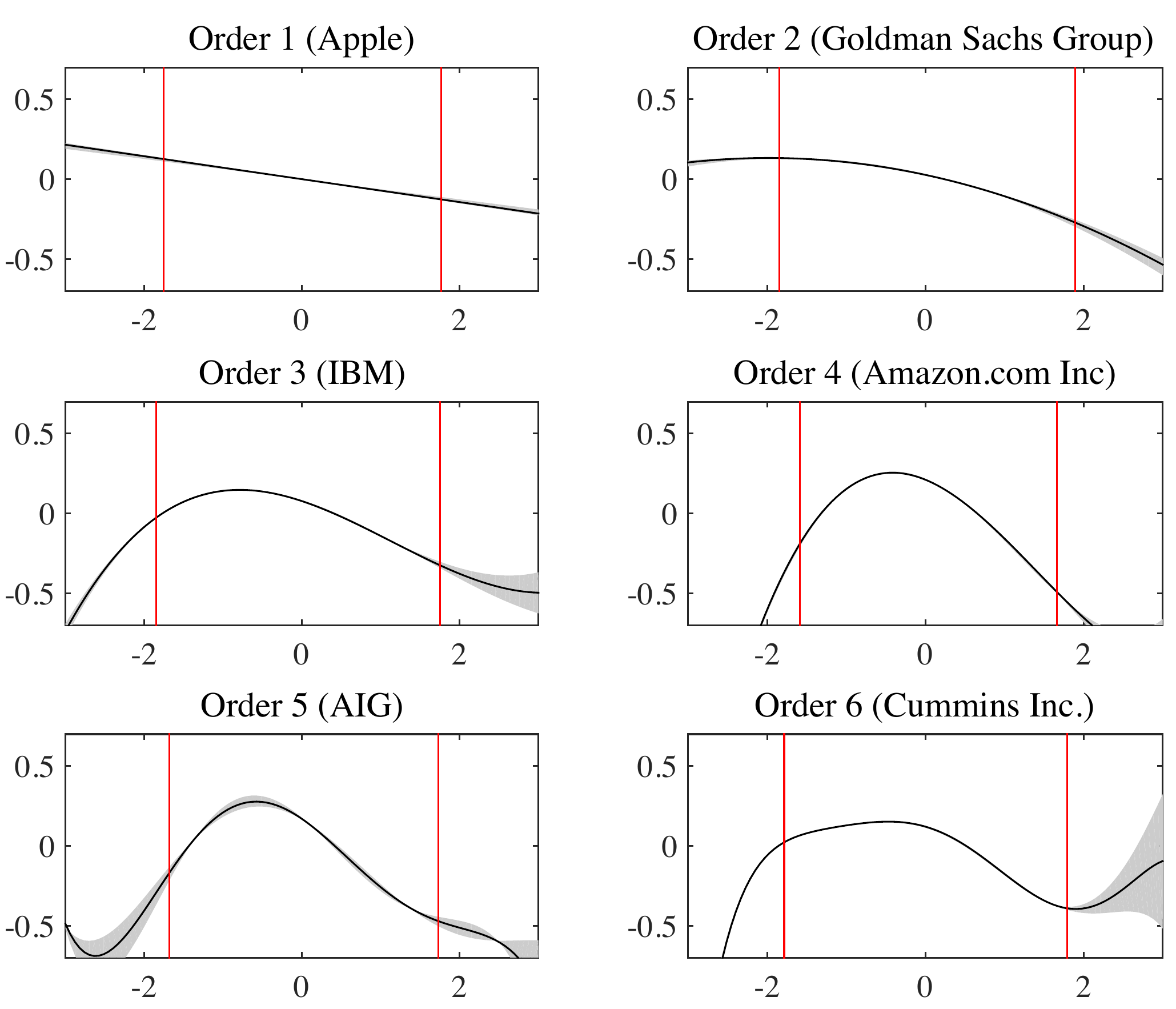}
\caption{Leverage functions of U.S. stocks (gray area: 95\% credible interval of the posterior distribution, vertical lines: 95\% interval of the observed returns). The x-axis is the shock to the return and the y-axis is the effect on volatility.}
\label{lev_sp}
\end{center}
\end{figure}

\paragraph{Examining leverage functions.} Table.~\ref{tab1} reports the number of positive/negative highest order coefficients, $\varphi_{k}$, in the Hermit polynomial of the leverage function and Figure.~\ref{lev_sp} exhibits the estimated leverage functions, i.e. news impact curves, for selected stocks that are optimal under that order, as an illustration.
Note that, for Figure.~\ref{lev_sp}, the stocks are selected arbitrarily and do not represent the whole population, though many stocks exhibit similar traits, as we can infer from Table.~\ref{tab1}.
In terms of the sign of the highest order coefficient, this informs us on the rough shape of the leverage function.
For example, the sign for the 1st-order leverage is negative for all stocks, agreeing with the consensus that leverage effects are a negative correlation.
For 2nd-order leverage functions, the sign of $\varphi_2$ determines whether the leverage function is concave ($\varphi_2<0$) or convex ($\varphi_2>0$). 
As we can see in Table.~\ref{tab1}, $\varphi_2$ is negative for most of the stocks, meaning that the leverage function is concave, similar to the 2nd-order leverage function in Figure.~\ref{lev_sp}. 
As for the estimated leverage functions in Figure.~\ref{lev_sp}, a concave leverage function has a mode somewhere around zero. As a result, a large shock, either positive or negative, will reduce the volatility, which is quite counterintuitive and a stark deviation from the conventional interpretation of the leverage effect.

The two key differences exhibited in the nonlinear leverage compared to the linear leverage are at zero and at the tail.
Firstly, since Hermite polynomials do not necessitate that the function be centered, we see that the assumption that an indifference in price equates in an indifference in volatility is false.
This suggests that stocks become more volatile after small, or no, movement in price, in either direction.
This may be caused by either investor sentiment to buy/sell stocks that are stagnate, intraday volatility not being accounted for, the model excessively shrinking its volatility estimate, or a combination of all.
From an economic perspective, we can infer that, at least for small fluctuations, the leverage effect does not come from financial leverage, but rather reflects investor sentiment and uncertainty.

Secondly, we observe that, although the shapes of the functions are different with order, they all roughly have a negative slope around zero similar to the 1st-order (linear) leverage function with the tail falling off to negative.
This result goes against the conventional interpretation of the leverage effect, as it implies that a large shock, either positive or negative, will reduce its volatility.
There are multiple possible interpretations as to why this might happen, and why it might better explain price movements in stocks.
One possible interpretation is that a sharp rise or fall in a stock price is a first moment phenomenon and not a second moment phenomenon. 
In other words, a stock price will drastically rise or fall in response to an unexpected change in the expected return of the stock, but not because of it's change in volatility. 
Another possible explanation is that, when a stock price rises or falls, the upward or downward trend tends to continue, i.e., the momentum effect, and this persistency reduces daily volatility. 
Though these possible interpretations are speculative, it is clear that these higher order forms better represent the data compared to the linear leverage effect and improve predictive performance.

\section{Conclusion \label{sec:conc}}

The leverage effect is a a critical factor in characterizing and forecasting volatility.
While observations from the market-- added with an accompanying sensible economic argument-- has brought on consensus among researchers and practitioners that this effect exists, empirical evidence from individual stocks were often contradictory, leading to poor performance.
In this paper, we have argued that this in not because the leverage effect does not exist, but because of our presumption of the effect, and how it is modeled, was wrong.
Under the stochastic volatility (SV) framework, we generalized the leverage effect, which has been presumed to be linear, to account for nonlinearity.
Applying the new model to stocks comprising the S\&P 500 or Nikkei 225, we have found that most stocks exhibit a nonlinear leverage effect, and that accounting for it drastically improves volatility forecasts.

Furthermore, this paper discussed new methods for estimating complex non-linear models using an extension of the particle learning algorithm.
With this new algorithm, we are now able to estimate and infer on models and parameters that were either too costly or too difficult using conventional methods, allowing researchers to flexibly extend current models and create new ones.
Additionally, the sequential nature of the model lends itself to real time applications of models that were deemed impractical before.

\bibliographystyle{asa}
\bibliography{reference} 

\begin{thebibliography}{39}
\newcommand{\enquote}[1]{``#1''}
\expandafter\ifx\csname natexlab\endcsname\relax\def\natexlab#1{#1}\fi

\bibitem[{Aastveit et~al.(2015)Aastveit, Ravazzolo, and van
  Dijk}]{Aastveit2015}
Aastveit, K.~A., Ravazzolo, F., and van Dijk, H.~K. (2015), \enquote{Combined
  density {N}owcasting in an uncertain economic environment,} \textit{Journal
  of Business \& Economic Statistics}, -, --, to appear.

\bibitem[{Bekaert and Wu(2000)}]{BW00}
Bekaert, G. and Wu, G. (2000), \enquote{Asymmetric volatility and risk in
  equity markets,} \textit{Review of Financial Studies}, 13, 1--42.

\bibitem[{Black(1976)}]{B76}
Black, F. (1976), \enquote{Studies of stock price volatility changes,}
  \textit{In: Proceedings of the 1976 Meetings of the American Statistical
  Association}, 171--181.

\bibitem[{Carvalho et~al.(2010)Carvalho, Johannes, Lopes, and Polson}]{CJLP10}
Carvalho, C.~M., Johannes, A.~M., Lopes, H.~F., and Polson, N.~G. (2010),
  \enquote{Particle learning and smoothing,} \textit{Statistical Science}, 25,
  88--106.

\bibitem[{Carvalho and Lopes(2007)}]{Carvalho2007}
Carvalho, C.~M. and Lopes, H.~F. (2007), \enquote{Simulation-based sequential
  analysis of Markov switching stochastic volatility models,}
  \textit{Computational Statistics \& Data Analysis}, 51, 4526--4542.

\bibitem[{Chib et~al.(2002)Chib, Nardari, and Shephard}]{Chib2002}
Chib, S., Nardari, F., and Shephard, N. (2002), \enquote{Markov chain Monte
  Carlo methods for stochastic volatility models,} \textit{Journal of
  Econometrics}, 108, 281--316.

\bibitem[{Christie(1982)}]{C82}
Christie, A.~A. (1982), \enquote{The stochastic behavior of common stock
  variances: Value, leverage and interest rate effects,} \textit{Journal of
  Financial Economics}, 10, 407--432.

\bibitem[{Creal(2012)}]{Creal2012}
Creal, D. (2012), \enquote{A survey of sequential Monte Carlo methods for
  economics and finance,} \textit{Econometric Reviews}, 31, 245--296.

\bibitem[{Dumas et~al.(1998)Dumas, Fleming, and Whaley}]{Dumas1998}
Dumas, B., Fleming, J., and Whaley, R. (1998), \enquote{Implied volatility
  functions: empirical tests,} \textit{Journal of Finance}, 53, 2059--2106.

\bibitem[{Fearnhead(2002)}]{Fearnhead02}
Fearnhead, P. (2002), \enquote{Markov chain Monte Carlo, sufficient statistics,
  and particle filters,} \textit{J. Comput. Graph. Statist.}, 11, 848--862.

\bibitem[{Fridman and Harris(1998)}]{Fridman1998}
Fridman, M. and Harris, L. (1998), \enquote{A maximum likelihood approach for
  non-Gaussian stochastic volatility models,} \textit{Journal of Business and
  Economics Statistics}, 16, 284--291.

\bibitem[{Gamerman and Lopes(2006)}]{Lopes2006}
Gamerman, D. and Lopes, H.~F. (2006), \textit{Markov Chain Monte Carlo:
  Stochastic Simulation for Bayesian Inference}, Chapman \& Hall/CRC Press.

\bibitem[{Geweke(1994)}]{Geweke1994}
Geweke, J. (1994), \enquote{Bayesian comparison of econometric models,}
  \textit{Working Paper, Federal Reserve Bank of Minneapolis Research
  Department}.

\bibitem[{Glosten et~al.(1993)Glosten, Jagannathan, and Runkle}]{Glosten1993}
Glosten, L., Jagannathan, R., and Runkle, D. (1993), \enquote{On the relation
  between the expected value and the volatility of the nominal excess return on
  stocks,} \textit{Journal of Finance}, 48, 1779--1801.

\bibitem[{Gordon et~al.(1993)Gordon, Salmond, and Smith}]{GSS93}
Gordon, N.~J., Salmond, D.~J., and Smith, A. F.~M. (1993), \enquote{Novel
  Approach to Nonlinear/Non-Gaussian Bayesian State Estimation,} \textit{IEEE
  Proceedings-F}, 140, 107--113.

\bibitem[{Hansen et~al.(2012)Hansen, Huang, and Shek}]{HHS11}
Hansen, P.~R., Huang, Z., and Shek, H.~H. (2012), \enquote{Realized GARCH: a
  Joint Model For Returns and Realized Mesures of Volatility,} \textit{Journal
  of Applied Econometrics}, 27, 877--906.

\bibitem[{Jacquier et~al.(1994)Jacquier, Polson, and Rossi}]{JPR94}
Jacquier, E., Polson, N.~G., and Rossi, P.~E. (1994), \enquote{Bayesian
  Analysis of Stochastic Volatility Models,} \textit{Journal of Business and
  Economic Statistics}, 12, 371--389.

\bibitem[{Jacquier et~al.(2004)Jacquier, Polson, and Rossi}]{JPR04}
--- (2004), \enquote{Bayesian analysis of stochastic volatility models with
  fat-tails and correlated errors,} \textit{Journal of Econometrics}, 122,
  185--212.

\bibitem[{Johannes and Polson(2008)}]{JP08}
Johannes, M. and Polson, N.~G. (2008), \enquote{Exact Particle Filtering and
  Learning,} \textit{Working paper}, Univ. Chicago Booth School of Business.

\bibitem[{Johannes et~al.(2008)Johannes, Polson, and Yae}]{JPY08}
Johannes, M., Polson, N.~G., and Yae, S.~M. (2008), \enquote{Non-linear
  Filtering and Learning,} \textit{Working paper}, Univ. Chicago Booth School
  of Business.

\bibitem[{Kim et~al.(1998)Kim, Shephard, and Chib}]{Kim1998}
Kim, S., Shephard, N., and Chib, S. (1998), \enquote{Stochastic volatility:
  Likelihood inference and comparison with ARCH models,} \textit{Review of
  Economic Studies}, 65, 361--393.

\bibitem[{Kitagawa(1996)}]{K96}
Kitagawa, G. (1996), \enquote{Monte Carlo Filter and Smoother for Non-Gaussian
  Nonlinear State Space Models,} \textit{Journal of Computational and Graphical
  Statistics}, 5, 1--25.

\bibitem[{Liu and West(2001)}]{LW01}
Liu, J. and West, M. (2001), \enquote{Combined Parameters and State Estimation
  in Simulation-based Filtering,} \textit{Sequential Monte Carlo Methods in
  Practice}, (A. Doucet, N. de Freitas and N. Gordon, eds.), 197--223.

\bibitem[{Lopes and Polson(2010)}]{Lopes2010}
Lopes, H.~F. and Polson, N.~G. (2010), \enquote{Bayesian inference for
  stochastic volatility modeling,} in \textit{Rethinking Risk Measurement and
  Reporting: Uncertainty, Bayesian Analysis and Expert Judgement}, ed.
  K.Booker, Risk Books, pp. 515--551.

\bibitem[{Lopes and Tsay(2011)}]{Lopes2011}
Lopes, H.~F. and Tsay, R.~S. (2011), \enquote{Particle Filters and Bayesian
  Inference in Financial Econometrics,} \textit{Journal of Forecasting}, 30,
  168--209.

\bibitem[{McAlinn and West(2017)}]{McAlinn2016}
McAlinn, K. and West, M. (2017), \enquote{Dynamic Bayesian predictive synthesis
  in time series forecasting,} \textit{Journal of Economietrics}, Forthcoming.

\bibitem[{Nakajima and Omori(2009)}]{Nakajima2009}
Nakajima, J. and Omori, Y. (2009), \enquote{Leverage, heavy-tails and
  correlated jumps in stochastic volatility models,} \textit{Computational
  Statistics \& Data Analysis}, 53, 2335--2353.

\bibitem[{Nakajima and Omori(2012)}]{Nakajima2012}
--- (2012), \enquote{Stochastic volatility model with leverage and
  asymmetrically heavy-tailed error using GH skew Student’s t-distribution,}
  \textit{Computational Statistics \& Data Analysis}, 56, 3690--3704.

\bibitem[{Nakajima and West(2013)}]{Nakajima2010}
Nakajima, J. and West, M. (2013), \enquote{Bayesian analysis of latent
  threshold dynamic models,} \textit{Journal of Business \& Economic
  Statistics}, 31, 151--164.

\bibitem[{Nelson(1991)}]{Nelson1991}
Nelson, D. (1991), \enquote{Conditional heteroskedasticity in asset pricing: A
  new approach,} \textit{Econometrica}, 59, 347--370.

\bibitem[{Omori et~al.(2007)Omori, Chib, Shephard, and Nakajima}]{OCSJ07}
Omori, Y., Chib, S., Shephard, N., and Nakajima, J. (2007), \enquote{Stochastic
  volatility with leverage: fast likelihood inference,} \textit{Journal of
  Econometrics}, 140, 425--449.

\bibitem[{Pitt and Shephard(1999)}]{PS99}
Pitt, M. and Shephard, N. (1999), \enquote{Filtering via Simulation: Auxiliary
  Particle Filters,} \textit{J. Amer. Statist. Assoc.}, 94, 590--599.

\bibitem[{Polson et~al.(2004)Polson, Stroud, and M{\"u}ller}]{Polson2004}
Polson, N.~G., Stroud, J.~R., and M{\"u}ller, P. (2004), \enquote{Practical
  Filtering for Stochastic Volatility Models,} in \textit{State Space and
  Unobserved Components Models}, eds. Harvey, A.~C., Koopman, S.~J., and
  Shephard, N., Cambridge University Press, pp. 236--247.

\bibitem[{Polson et~al.(2008)Polson, Stroud, and M{\"u}ller}]{PSM08}
--- (2008), \enquote{Practical Filtering with Sequential Parameter Learning,}
  \textit{J. Roy. Statist. Soc. Ser. B}, 70, 413--428.

\bibitem[{Shirota et~al.(2014)Shirota, Hizu, and Omori}]{Shirota2014}
Shirota, S., Hizu, T., and Omori, Y. (2014), \enquote{Realized stochastic
  volatility with leverage and long memory,} \textit{Computational Statistics
  \& Data Analysis}, 76, 618--641.

\bibitem[{Storvik(2002)}]{Storvik02}
Storvik, G. (2002), \enquote{Particle Filters in State Space Models with the
  Presence of Unknown Static Parameters,} \textit{IEEE Trans. Signal Process},
  50, 281--289.

\bibitem[{Takahashi et~al.(2013)Takahashi, Omori, and Watanabe}]{Takahashi2013}
Takahashi, M., Omori, Y., and Watanabe, T. (2013), \enquote{News impact curve
  for stochastic volatility models,} \textit{Economics Letters}, 120, 130--134.

\bibitem[{West and Harrison(1997)}]{WH97}
West, M. and Harrison, J. (1997), \textit{Bayesian Forecasting and Dynamic
  Models}, Springer, 2nd ed.

\bibitem[{Yu(2005)}]{J05}
Yu, J. (2005), \enquote{On leverage in a stochastic volatility model,}
  \textit{Journal of Econometrics}, 127, 165--178.

\end{thebibliography}

\newpage

\begin{center}
{\Large Why Don't Stocks Exhibit Persistent Leverage Effects?} 

\bigskip

\bigskip
{\Large  Tables and Figures} 
\bigskip\bigskip
\end{center}

 \begin{table}[tbh]
\begin{center}
\begin{tabular}{l|ccc}
&$\mu$ & $\beta$ & $\tau$ \\
\hline 
True value&-0.026 & 0.970 & 0.150 \\
MCMC& -0.026 & 0.968 & 0.162\\
     & [-0.031 -0.021] & [0.951 0.980] & [0.130 0.205] \\
PL   & -0.034 & 0.961 & 0.174 \\
    & [-0.049 -0.023] & [0.947 0.972] & [0.145 0.199] \\
PLAV  & -0.030 & 0.965 & 0.163 \\
     & [-0.042 -0.020] & [0.952 0.975] & [0.146 0.191] \\
\end{tabular}
\end{center}
\caption{Posterior mean and credible interval (in brackets) of the parameters. Data simulated using the assigned true values in a no-leverage SV model.}
\label{pmsv}
\end{table}

\begin{table}[tbh]
\begin{center}
\begin{tabular}{l|cccc}
&$\mu$ & $\beta$ & $\varphi$&$\omega$ \\
\hline 
True value & -0.026 & 0.970 & -0.045 & 0.143 \\
MCMC & -0.020 & 0.978 & -0.014 & 0.132 \\
     & [-0.024 -0.015] & [0.967 0.987] & [-0.025 0.003] & [0.106 0.161]\\
PLAV  & -0.023 & 0.974 & -0.038 & 0.144 \\
    & [-0.032 -0.014] & [0.966 0.982] & [-0.047 -0.025] & [0.133 0.153]\\
\end{tabular}
\end{center}
\caption{Posterior mean and credible interval (in brackets) of the parameters. Data simulated using the assigned true values in a linear-leverage SV model.}
\label{pmlsv}
\end{table}

\begin{table}[tbh]
\begin{center}
\begin{tabular}{c|rr}
Algorithm & No Leverage (min.) & Linear Leverage (min.) \\
\hline
MCMC & 2.4  & 737.6 \\
PL   & 66.8 &  --   \\
PLAV  & 64.9 & 81.6 \\
\end{tabular}
\end{center}
\caption{Execution time of the three algorithms.}
\label{timesv}
\end{table} 

\begin{table}[ht!]
\begin{center}
\renewcommand{\tabcolsep}{5pt}
\begin{tabular}{ccccc}
Order of & \multicolumn{2}{c}{S\&P500} & \multicolumn{2}{c}{Nikkei225} \\
\cline{2-5}
Leverage & Positive & Negative & Positive & Negative \\
\hline 
1 & 0 & 50 & 0 & 32 \\
2 & 3 & 160 & 10 & 60  \\
3 & 58 & 28 & 18 & 15  \\
4 & 15 & 29 & 9 & 7  \\
5 & 9 & 11 & 5 & 3  \\
6 & 12 & 5 & 3 & 5  \\
\hline
\end{tabular}
\caption{The number of positive/negative highest order coefficients in the leverage function for each optimal leverage order selected using the predictive marginal likelihood.}
\label{tab1}
\end{center}
\end{table}

\end{document}